# End-to-end evaluation of research organizations

## 1. Introduction

Abramo and D'Angelo (2014, 2016) have recently looked at the issue of measuring performance and productivity of research organizations and the role that size-independent citation indicators play in this. Recently, Savithri and Prathap (2015) compared the research performance of leading higher education institutions in India and China using an end-to-end bibliometric performance analysis procedure with data from the 2014 release of the SCImago Institutions Rankings (SIR). Six primary and secondary bibliometric indicators were used to summarize the chain of activity: input–output–excellence–outcome–productivity. Principal component analysis indicated that the primary indicators are orthogonal and represent size-dependent quantity and size-independent quality/productivity dimensions respectively. Using this insight two-dimensional maps can be used to visualize the results. It was clear that to arrive at meaningful summary statistical indicators for performance and productivity, both size-dependent and size-independent indicators play a key role and that the bibliometric core of the chain (measuring output or outcome) must be separated from the econometric part of the chain (the outcome or output to input ratios) .

In the discussion below, we first introduce the role of size-dependent and size-independentt indicators in the bibliometric part of the evaluation chain. We show that performance can then be evaluated at various levels, namely a zeroth-order, a first-order or even a second-order. To complete the evaluation chain, we take up the econometric part where efficiency of the research production process is represented in terms of output and outcome productivities.

## 2. Size-dependent and size-independent indicators

An evocative analogy for understanding the relationship of size-dependent to size-independent factors in all measurement is Archimedes' discovery of the concept of density. The density $\rho$ is a size-independent term that allows the weight $W$ to be computed from the volume $V$, which is the primary size-dependent term. Note that now, $W$ combines both size-dependent and size-independent terms into a meaningful composite secondary indicator. The bibliometric parallel for this are $P$, the number of publications and $C$ the number of citations in a portfolio of publications. Thus, if $P$ is taken as the primary bibliometric indicator of size, then $C$ becomes a secondary and composite bibliometric indicator of performance. Impact, which is represented by $i = C/P$, is a natural candidate for a size-independent proxy for the quality of the portfolio. Of course at this stage, we assume that all publications are in the same discipline and from a coeval window so that normalization is not an issue. Normalization is only an additional detail that can be rationally worked out (Ruiz-Castillo & Waltman 2015).

## 3. Zeroth-, first- and second-order indicators

If $C$ is thought of as a first-order indicator of performance, then it is possible to bring in the idea of an higher-order energy-like term $X = iC = i^2P$, as another indicator of bibliometric performance. Thus, $C$ combines impact $i$ and output $P$ by weighting each publication with its citation impact. The I3 indicator (Leydesdorff & Bornmann, 2011) combines normalized impact and output and is therefore a first-order indicator of performance. The exergy indicator of Prathap (2011) is a second-order indicator of performance. $P$, standing alone, is then a zeroth-order indicator of performance. Thus all three, $P$, $C$ and $X$ are valid measures of output or outcome depending on the extent to which one wants to give weightage to the quality proxy, in this case, the impact $i$.

## 4. The econometric part of the chain

Let us now come to the econometric part of the chain. We need a meaningful measure of input as this is crucial to the calculation of the research efficiency or productivity of any research-intensive unit. In 2014, SIR introduced a new feature that makes end-to-end evaluation from input to output possible. This is called the scientific talent pool (STP), which gives the number of authors from an institution who have participated in the total publication output of that institution during that particular period of time. Savithri and Prathap (2015) used this indicator as a reasonable proxy of the input at the beginning of the chain that performs scientific research activity.

To the best of this author's knowledge, Hendrix (2008) was one of the earliest to evaluate institutional-level performance of research units by intelligently classifying and clustering various bibliometric indicators using Principal Component Analysis (PCA). The variables clustered neatly into three distinct groups: the first cluster comprise size-dependent input and output terms, namely the total number of faculty (input), total number of papers (output), and total number of citations (outcome). The second factor comprised size-independent terms that reflect the impact of a researcher, average number of citations per article, etc. and can be interpreted as a quality or excellence dimension. The third group, also influenced heavily by size-independent terms, describes research productivity and impact at the individual level, like the number of papers and number of citations per faculty member.

Savithri and Prathap (2015) used the PCA approach to show that with only five variables, two components suffice to account for most of the common variance. These are the size-dependent quantity indicators and the size-independent quality and productivity indicators, which are clearly orthogonal to the former. This allowed representation and visualization of the primary and secondary data as two-dimensional maps. Thus for an end-to-end evaluation, size-dependent and size-independent indicators play a very critical role.

## 4. A simple example and discussion of the results

In this paper, we represent the indicators needed for the complete end-to-end chain as shown in Table 1. Using this we rework the simple example in Abramo and Angelo (2016). Table 2 takes the case of two universities of the same size (say 100 Full time Equivalent Researchers or FTERs), resources and research fields. Unit A publishes 100 articles earning 1000 citations (i.e. impact of 10 citations per article). Unit B publishes 200 articles, and gathers a total of 1500 citations (i.e. average impact of 7.5 citations per article). The last column of Table 2 shows the efficiency or effectiveness advantage of B over A using the Mean Normalized Citation Score (MNCS) approach. Since performance is a multi-dimensional construct, we have different results - A is better than B on quality alone, but on output or outcome productivities, depending on the choice of order of indicator, the advantages change. The exercise can be repeated using the Highly Cited Articles (HCA) approach. Unit A has 10 HCAs while Unit B has 15 HCAs as shown in Table 3 and there is no change in the results.

## 5. Conclusions

End-to-end research evaluation needs to separate out the bibliometric part of the chain from the econometric part. Both size-dependent and size-independent terms play a crucial role to combine quantity and quality (impact) in a meaningful way. Output or outcome at the bibliometric level can be measured using zeroth, first or second-order composite indicators, and the productivity terms follow accordingly using the input to output or outcome factors.

## Acknowledgements

The author is thankful to Dr Kuncheria P Isaac, Vice Chancellor of the A P J Abdul Kalam Technological University and to Shri P K Asokan of the Vidya International Charitable Trust for their constant encouragement and support and for making available the facilities and resources of the university to conduct this research.

Gangan Prathap
*Vidya Academy of Science and Technology, Thrissur, Kerala, India 680501*
*and*
*A P J Abdul Kalam Technological University, Thiruvananthapuram, Kerala, India 695016*
*E-mail addresses*: gangan@vidyaacademy.ac.in and gangan_prathap@hotmail.com

Table 1. The primary indicators or variables and the derived indicators for the end-to-end chain.

| Indicator or variable | Description | Size dependence | Formula |
|---|---|---|---|
| **S** | FTER | Dependent | S |
| **P** | Output | Dependent | P |
| **i** | Excellence | Independent | C/P |
| **C** | 1st order outcome | Composite | C |
| **X** | 2nd order outcome | Composite | iC |
| **P/S** | Output productivity | Independent | P/S |
| **C/S** | 1st order outcome productivity | Independent | C/S |
| **X/S** | 2nd order outcome productivity | Independent | iC/S |

Table 2. Comparison of two universities from Abramo and Angelo (2016) using the MNCS approach.

| Indicator or variable | Unit A | Unit B | Percentage advantage |
|---|---|---|---|
| S | 100 | 100 | 0 |
| P | 100 | 200 | 100 |
| i | 10 | 7.5 | -25 |
| C | 1000 | 1500 | 50 |
| X | 10000 | 11250 | 12.5 |
| P/S | 1 | 2 | 100 |
| C/S | 10 | 15 | 50 |
| X/S | 100 | 112.5 | 12.5 |

Table 3. Comparison of two universities from Abramo and Angelo (2016) using the HCA approach.

| Indicator or variable | Unit A | Unit B | Percentage advantage |
|---|---|---|---|
| S | 100 | 100 | 0 |
| P | 100 | 200 | 100 |
| i | 0.1 | 0.075 | -25 |
| HCA | 10 | 15 | 50 |
| X | 1 | 1.125 | 12.5 |
| P/S | 1 | 2 | 100 |
| C/S | 0.1 | 0.15 | 50 |
| X/S | 0.01 | 0.01125 | 12.5 |